# Performance of a Nonempirical Density Functional on Molecules and Hydrogen-Bonded Complexes

Yuxiang Mo[1], Guocai Tian[1,2], Roberto Car[3], Viktor N. Staroverov[4], Gustavo E. Scuseria[5], and Jianmin Tao[1,*]

[1]Department of Physics, Temple University, Philadelphia, Pennsylvania 19122, USA
[2]State Key Laboratory of Complex Nonferrous Metal Resources Clean Utilization, Kunming University of Science and Technology, Kunming 650093, China
[3]Department of Chemistry, Princeton University, Princeton, New Jersey 08544, USA
[4]Department of Chemistry, The University of Western Ontario, London, Ontario N6A 5B7, Canada
[5]Department of Chemistry, Rice University, Houston, Texas 77005, USA

### Abstract

Recently, Tao and Mo (TM) derived a meta-generalized gradient approximation functional based on a model exchange-correlation hole. In this work, the performance of this functional is assessed on standard test sets, using the 6-311++G(3$df$,3$pd$) basis set. These test sets include 223 G3/99 enthalpies of formation, 99 atomization energies, 76 barrier heights, 58 electron affinities, 8 proton affinities, 96 bond lengths, 82 harmonic vibrational frequencies, 10 hydrogen-bonded molecular complexes, and 22 atomic excitation energies. Our calculations show that the TM functional can achieve high accuracy for most properties considered, relative to the LSDA, PBE, and TPSS functionals. In particular, it yields the best accuracy for proton affinities, harmonic vibrational frequencies, hydrogen-bonded dissociation energies and bond lengths, and atomic excitation energies.

## 1. INTRODUCTION

Kohn-Sham density functional theory (DFT)[1] provides an efficient description of the electronic structure of molecules and solids. In this theory, only the exchange-correlation energy component accounting for all many-body effects must be approximated as a functional of the electron density. Owing to the rapid development of exchange-correlation density functional approximations,[2–25] DFT has achieved a high degree of sophistication and become a standard technique of electronic structure calculations. However, despite considerable progress in the development of density functional approximations, there remains a strong demand for new density functionals with higher accuracy and wider applicability.[26,27]

Depending on the type of their ingredients, density functionals can be divided into two broad categories: semilocal and nonlocal. Semilocal functionals employ local or semilocal information, such as the electron density, density gradient, and the Kohn-Sham kinetic energy density, to calculate the exchange-correlation energy, while nonlocal functionals [23,28–31] make use of additional information beyond that of semilocal DFT, such as the exact exchange energy density. Nonlocal functionals provide more accurate description than semilocal approximations for problems in which nonlocality is important (e.g., band gaps, excitation energy, charge transfer, and reaction barriers), but they are computationally more expensive and more difficult to develop and implement. Semilocal DFT can be further divided into three sub-categories: local spin-density approximation (LSDA) [32] which uses the local spin-densities as inputs, generalized-gradient approximation (GGA) [3,6,33-37] which takes the spin-density gradients as additional inputs, and meta-GGA [9,20,38] with the kinetic energy densities as additional inputs. The functional form of GGAs is quite restrictive, but the form of meta-GGAs is more flexible. This flexibility allows meta-GGAs to satisfy more exact constraints and thus leads to improvement over GGA in accuracy and applicability. For example, a GGA can satisfy the exact second-order gradient expansion,[18] but only a meta-GGA [9,20,24,39] can simultaneously recover the correct fourth-order term. A GGA cannot be one-electron self-interaction free, but a meta-GGA correlation can.

Recently, two of the present authors (JT and YM) derived a meta-GGA functional based on an exchange-correlation hole, referred to hereafter as the Tao-Mo (TM) functional.[39] The exchange part of the hole was obtained from the density matrix expansion under an appropriate coordinate transformation, while the correlation part was taken from the Constantin-Perdew-Tao correlation hole[40] with a modification aiming to improve the low-density or strong-interaction limit of the correlation energy. This functional follows the non-empiricism philosophy of the widely-used Tao-Perdew-Staroverov-Scuseria (TPSS) density functional,[9] without relying on any empirical fitting, except for the exchange energy of the ground-state H atom.

In this paper, we present a comprehensive evaluation of the performance of the TM functional on a variety of properties of molecules and hydrogen-bonded complexes. We show that TM can achieve high accuracy for most properties considered here, among non-empirical density functionals proposed in recent years. For some properties and hydrogen-bonded complexes, it even gives the smallest error, in comparison with the benchmark data reported in the literature. Our assessment suggests that TM functional is a promising tool for electronic structure calculations.

## 2. COMPUTATIONAL METHOD

Because the exchange and correlation parts of a density functional have different coordinate [41] and spin [42,43] scaling properties, they are usually approximated separately. In the development of an exchange functional, one only needs to consider spin-unpolarized densities. The spin-polarized form is then obtained by the exact spin-scaling relationship,[42] $E_x[n_\uparrow, n_\downarrow] = E_x[2n_\uparrow]/2 + E_x[2n_\downarrow]/2$. For the correlation part, the exact spin-dependence is known only in the high-density limit.[44] Therefore, in the development of a correlation functional, one has to consider its spin-dependence for any spin polarization.

For spin-unpolarized densities, the exchange part of the TM meta-GGA functional [39] takes the form

$$E_x[n] = \int d^3r \, n \, \epsilon_x^{\text{unif}}(n) F_x(n, \nabla n, \tau), \tag{1}$$

where $n$ is the electron density, $\epsilon_x^{\text{unif}}(n)$ is the exchange energy per electron of the uniform electron gas given by $\epsilon_x^{\text{unif}}(n) = -3(3\pi^2 n)^{1/3}/4\pi$, $\tau(\mathbf{r}) = \sum_{i=1}^{N/2} |\nabla \phi_i(\mathbf{r})|^2$ is the total kinetic energy density, and $F_x$ is the enhancement factor. The inhomogeneity effects enter the meta-GGA functional via the enhancement factor, which was derived from the exchange hole via the density matrix expansion (DME) and finally corrected to satisfy the fourth-order gradient expansion of the exchange energy for the slowly varying density.[39] The slowly varying correction may not be so significant for molecular systems, but it is important for solids and surfaces, because the typical valence electron density of bulk solids is slowly varying. The TM exchange enhancement factor is expressed as

$$F_x = w F_x^{\text{DME}} + (1-w) F_x^{\text{SC}}, \tag{2}$$

where $w$ is the weight factor given by

$$w = \frac{(\tau_W/\tau)^2 + 3(\tau_W/\tau)^3}{\left[1+(\tau_W/\tau)^3\right]^2}. \tag{3}$$

$F_x^{SC}$ represents the slowly varying correction (SC)

$$F_x^{SC} = \left\{1+10\left[\left(\frac{10}{81}+\frac{50}{729}p\right)p+\frac{146}{2025}\tilde{q}^2\right.\right.$$

$$\left.\left.-\frac{73}{405}\tilde{q}\frac{3}{5}\left(\frac{\tau_W}{\tau}\right)\left(1-\frac{\tau_W}{\tau}\right)\right]+0\cdot p^2\right\}^{1/10}, \tag{4}$$

where $\tilde{q} = 3\tau/2k_F^2 n - 9/20 - p/12$, $p = s^2 = (|\nabla n|/2k_F n)^2$, and $\tau_W = |\nabla n|^2/8n$ is the von Weizsäcker kinetic energy density. $F_x^{DME}$ is the exchange enhancement factor obtained from the DME. It is given by

$$F_x^{DME} = \frac{1}{f^2} + \frac{7}{9f^4}\left\{1+\frac{595}{54}(2\lambda-1)^2 p\right.$$

$$\left.-\frac{1}{\tau^{unif}}\left[\tau-3\left(\lambda^2-\lambda+\frac{1}{2}\right)\left(\tau-\tau^{unif}-\frac{1}{72}\frac{|\nabla n|^2}{n}\right)\right]\right\}, \tag{5}$$

where $\tau^{unif} = 3(3\pi^2)^{2/3}n^{5/3}/10$ is the Thomas-Fermi kinetic energy density, $f = \left[1+10(70y/27)+\beta y^2\right]^{1/10}$, $y = (2\lambda-1)^2 p$, $\lambda = 0.6866$, and $\beta = 79.873$. In iso-orbital regions (e.g., core and density tail regions), $w \to 1$ so that $F_x \to F_x^{DME}$, while in the slowly varying density limit, $w \to 0$ and $F_x \to F_x^{SC}$. Therefore, we may interpret TM exchange as an interpolation between rapidly varying and slowly varying densities, similar to the TPSS exchange.[9]

The correlation part of the TM functional was developed by modifying the TPSS correlation approximation in the low-density (strong-interaction) limit. It takes the same form as that of TPSS, but replaces $C(\zeta,\xi)$ of Eq. (14) of Ref. 9 with the simpler form:

$$C(\zeta,\xi) = \frac{0.1\zeta^2 + 0.32\zeta^4}{\left\{1+\xi^2\left[(1+\zeta)^{-4/3}+(1-\zeta)^{-4/3}\right]/2\right\}^4} \quad (6)$$

where $\zeta = (n_\uparrow - n_\downarrow)/n$ is the relative spin polarization and $\xi = |\nabla\zeta|/2k_F$.[35,45] In the low-density limit, the exchange-correlation energy should become spin-independent, because two charged particles far apart from each other interact via the Coulomb interaction, regardless of whether they are bosons or fermions.[46] For example, in the dissociation limit of the $H_2$ molecule, each H atom can be spin-up or spin-down, without changing the total energy of the dissociated molecule. This limit was used to construct the TPSS correlation functional. It has also been recently employed to improve the TPSS correlation for the one-electron Gaussian density, leading to the TM correlation functional (see Fig. 2 of Ref. 39 for comparison of TM and TPSS). Like the TPSS correlation, the TM correlation satisfies two other exact constraints: (i) It recovers the slowly varying gradient expansion,[44] and (ii) it is one-electron self-interaction-free.

A nice feature of the TM functional is that the underlying exchange-correlation hole is known. The exchange part of the hole was derived from the DME, while the correlation part takes the form proposed by Constantin, Perdew, and Tao,[40] with the TPSS correlation energy density replaced by the TM correlation energy density. (The modification of the TPSS correlation energy is equivalent to the modification of the TPSS correlation hole, because the latter can be reverse-engineered[40] from the former.)

In the present work, we focus on the performance of the TM functional on energetic and structural properties of molecules. The tested properties include standard enthalpies of formation, atomization energies, reaction barrier heights, electron affinities, proton affinities, bond lengths, vibrational frequencies, H-bond dissociation energies and bond lengths, and atomic excitation energies. In order for the assessment to be reliable, we adopted the large basis set 6-311++G(3*df*,3*pd*) for most of our calculations. All integrals were evaluated on ultrafine grids (Grid=UltraFine). All molecular geometry optimizations were performed with the Opt=Tight option. The TM functional was implemented by modifying the Gaussian 09 program.[47] We use the mean error (ME), the mean absolute error (MAE), and the largest individual deviation to characterize and compare the accuracy of various density functionals. Calculated properties of individual species are available in the supplementary material.[48]

## 3. RESULTS AND DISCUSSION

### 3.1 Thermochemical Properties

In the present work, we assess the accuracy of the TM functional on thermochemical properties of the G3/99 and W4-08 test sets. The G3/99 test set was originally introduced by Curtiss and co-workers in their Gaussian-1,[49] Gaussian-2,[50] and Gaussian-3[51] theories for calculation of molecular energies and comparison with experimental data. It includes 223 standard enthalpies of formation (55 original G2 molecules,[50] 93 additional molecules,[52] and 75 larger organic molecules and inorganic compounds[53]), 58 electron affinities, and 8 proton affinities. Only the first- and second-row elements (Z<18) are represented. The G3/99 set has been widely used for the assessment and calibration of new theoretical methods. In addition, the W4-08 test set [54] of 99 small neutral molecules is used to evaluate the TM functional on atomization energies.

### 3.1.1 Standard enthalpies of formation

The standard enthalpy of formation is defined as the enthalpy change during the chemical reaction in which one mole of the compound is formed from its constituent elements, with all substances in their standard states at 1 atm (1 atm = 101.3 kPa). Standard enthalpies of formation at 298 K ($\Delta_f H^o_{298}$) were obtained from total atomic and molecular energies using the experimental atomic data and methodology described by Curtiss *et al.*[53,55] In order to make direct comparison of the TM functional with other DFT methods reported in the literature, in this work we adopt the procedure of Staroverov *et al.*[19] which uses the equilibrium B3LYP/6-31G(2*df*,*p*) geometries in combination with the B3LYP/6-31G(2*df*,*p*) zero-point energies (ZPE) and thermal corrections obtained with a frequency scale factor of 0.9854. Total electronic energies are calculated for those geometries using the much larger basis set 6-311++G(3*df*,3*pd*).

As shown in Table I (see Tables S1 and S2 for molecule-specific data), the TM functional is more accurate for standard enthalpies of formation than many other approximations, but it is less accurate than the VSXC, TPSS, OLYP, HCTH, and hybrid functionals. Similar to other functionals, but unlike TPSS, the error of the TM functional increases with increasing molecular size (from G2 to G3). However, the rate of this error increase is the smallest for TM, compared to other functionals. The largest error occurs for molecules containing reference atoms with a relatively large spin polarization such as O, S, N, Si, F, and Cl, as in other methods except TPSS.

TABLE I. Summary of deviations from experiment of the calculated $\Delta_f H^o_{298}$ for the G3/99 test set. Results of other functionals are taken from Ref. 19. All values are in kcal/mol. For non-hybrid functionals, the smallest and largest MAEs are in bold blue and red, respectively.

| Method | G2 set/148 | | | | G3 set/75 | | | | G3/99 223 | |
|---|---|---|---|---|---|---|---|---|---|---|
| | ME | MAE | Max(+) | Max(-) | ME | MAE | Max(+) | Max(-) | ME | MAE |
| Non-hybrid | | | | | | | | | | |
| LSDA | -83.7 | **83.7** | 0.4 (Li$_2$) | -207.7 (C$_6$H$_6$) | -197.1 | **197.1** | None | -347.5 (azulene) | -121.9 | **121.9** |
| BLYP | -0.6 | 7.3 | 24.2 (SiCl$_4$) | -28.1 (NO$_2$) | 12.4 | 13.9 | 41.0 (C$_8$H$_{18}$) | -11.0 (C$_4$H$_4$N$_2$) | 3.8 | 9.5 |
| BPW91 | -5.4 | 8.0 | 16.5 (SiF$_4$) | -32.4 (NO$_2$) | -5.0 | 11.1 | 22.4 [Si(CH$_3$)$_4$] | -28.0 (azulene) | -5.3 | 9.0 |
| BP86 | -19.9 | 20.1 | 7.1 (SiF$_4$) | -48.7 (C$_5$H$_5$N) | -38.6 | 38.6 | None | -72.7 (azulene) | -26.2 | 26.3 |
| PW91 | -17.2 | 17.7 | 7.5 (Si$_2$H$_6$) | -52.7 (C$_2$F$_4$) | -35.3 | 35.3 | None | -81.1 (azulene) | -23.3 | 23.6 |
| PBE | -16.1 | 16.9 | 10.8 (Si$_2$H$_6$) | -50.5 (C$_3$F$_4$) | -32.8 | 32.8 | None | -79.7 (azulene) | -21.7 | 22.2 |
| HCTH | -0.6 | 5.6 | 16.5 (SiCl$_4$) | -28.0 (C$_3$F$_4$) | 6.4 | 10.2 | 27.5 [Si(CH$_3$)$_4$] | -22.2 (C$_2$F$_6$) | 1.7 | 7.2 |
| OLYP | -1.9 | 4.8 | 27.0 (SiF$_4$) | -23.5 (NO$_2$) | 6.4 | 7.9 | 20.9 [Si(CH$_3$)$_4$] | -11.0 (CF$_3$) | 0.9 | 5.9 |
| VSXC | -0.5 | **2.8** | 8.2 (N$_2$H$_4$) | -11.5 (CS$_2$) | 1.97 | **4.7** | 12.0 (C$_8$H$_{18}$) | -8.7 (C$_6$H$_5$) | 0.3 | **3.5** |
| TPSS | -5.2 | 6.0 | 16.2 (SiF$_4$) | -22.9 (ClF$_3$) | -5.2 | 5.5 | 7.5 (PF$_5$) | -12.8 (S$_2$Cl$_2$) | -5.2 | 5.8 |
| TM | -2.6 | 6.8 | 23.4 (SiF$_4$) | -20.7 (NF$_3$) | -2.8 | 9.6 | 12.8 (PF$_5$) | -16.4 (nitro-s-butane) | -2.6 | 7.8 |
| Hybrid | | | | | | | | | | |
| B3LYP | 1.1 | 3.1 | 20.1 (SiF$_4$) | -8.1 (BeH) | 8.2 | 8.4 | 20.8 (SF$_6$) | -4.9 (C$_4$H$_4$N$_2$) | 3.5 | 4.9 |
| B3PW91 | -1.4 | 3.4 | 21.6 (SiF$_4$) | -12.8 (C$_2$F$_4$) | -2.5 | 4.9 | 17.0 (PF$_5$) | -17.0 (naphthalene) | -1.8 | 3.9 |
| B3P86 | -17.9 | 18.2 | 7.5 (SiF$_4$) | -48.1 (C$_5$H$_8$) | -41.9 | 41.9 | None | -79.2 (C$_8$H$_{18}$) | -26.0 | 26.1 |
| PBE0 | -2.4 | 4.9 | 21.3 (SiF$_4$) | -19.8 (C$_5$H$_5$N) | -9.3 | 10.2 | 14.5 (PF$_5$) | -35.6 (naphthalene) | -4.7 | 6.7 |
| TPSSh | -1.4 | 4.2 | 22.0 (SiF$_4$) | -18.0 (Si$_2$H$_6$) | 0.2 | 3.3 | 16.2 (PF$_5$) | -6.6 (C$_8$H$_{18}$) | -0.9 | 3.9 |

### 3.1.2 Atomization energies

The atomization energy of a molecule is defined as the difference between the total energies of the molecule and the free constituent atoms, all at 0 K. In the present work, the atomization energies were evaluated for the W4-08 test set,[54] which includes 99 small molecules. The equilibrium geometries of all the molecules in this test set and electronic energies (not including ZPE) for those geometries were obtained using the 6-311++G(3*df*,3*pd*) basis set. Listed in Table II are the statistical errors on atomization energies. The ME of the TM functional for W4-08 atomization energies is +3.11 kcal/mol, suggesting an overestimation trend. The largest MEs are for $F_2O_2$ (+32.67 kcal/mol) and $AlF_3$ (-19.00 kcal/mol). For all the density functionals considered, the largest errors are observed for molecules containing reference atoms with a relatively large spin polarization such as B, Al, O, N, Si, F, and Cl. The TM functional has a mean absolute error of 7.43 kcal/mol, which is larger than that of M06L (MAE = 4.56 kcal/mol), OLYP (MAE = 5.33 kcal/mol), and TPSS (MAE = 5.24 kcal/mol), but much smaller than those of PBE (MAE = 12.97 kcal/mol) and LSDA (MAE = 47.43 kcal/mol). We also analyzed separately the 53 G2 molecules included in the W4-08 set. The MAEs for the G2 subset are marginally smaller than the corresponding MAEs of the total W4-08 set for every functional. However, the MAEs for the total W4-08 test set are significantly smaller than those for the G3 test set containing 75 larger molecules. Part of the reason for this disparity is that dispersion interactions between atoms in a molecule are more important for the larger G3 molecules than for the W4-08 set, but conventional density functionals cannot entirely capture these interactions. (Standard DFT methods also miss nonlocal long-range van der Waals interactions and for that reason have difficulty describing intermolecular forces.[56–63])

Apart from the inadequate description of dispersion interaction, there are two other potential sources of error. First, the atomization energy of a molecule depends on the accuracy of atomic energies, which could be problematic.[64] In most cases, the electron density is spin-unpolarized in a molecule, but spin-polarized in the constituent atoms. Although the spin-dependence in the exchange part of a density functional is exact due to the simple spin scaling relationship, the spin-dependence of correlation energy is not, meaning that the spin-dependence of atomic energies may be less accurate than that of molecular energies. Second, semilocal functionals make relatively large errors for molecules with electrons occupying antibonding orbitals, where the electron density is rapidly varying.

TABLE II. Summary of deviations of the calculated atomization energies from CCSD(T)[54] for the W4-08 test set. The second and third columns are MEs and MAEs for the 53 molecules in the W4-08 set which also belong to the G2 set. The fourth and fifth columns are MEs and MAEs for the entire W4-08 set. Results of other functionals are taken from Ref. 19. All values are in kcal/mol. For non-hybrid functionals, the smallest and largest MAEs are in bold blue and red, respectively.

|  | 53 G2 molecules | | Entire W4-08 Set | | | |
| --- | --- | --- | --- | --- | --- | --- |
| Method | ME | MAE | ME | MAE | Max (+) | Max (−) |
| Non-hybrid | | | | | | |
| LSDA | 46.35 | **46.35** | 47.43 | **47.43** | 123.30 ($F_2O_2$) | |
| BLYP | 3.84 | 6.96 | 4.34 | 6.98 | 37.70 ($F_2O_2$) | −23.19 ($AlCl_3$) |
| BP86 | 9.72 | 10.46 | 10.28 | 11.03 | 45.45 ($F_2O_2$) | −12.00 ($AlCl_3$) |
| PBE | 10.02 | 11.83 | 11.40 | 12.97 | 53.51 ($F_2O_2$) | −15.40 ($Si_2H_6$) |
| OLYP | 2.20 | 4.89 | 2.61 | 5.33 | 29.94 ($F_2O_2$) | −19.46 ($AlCl_3$) |
| TPSS | 2.39 | 4.54 | 3.14 | 5.24 | 25.06 ($F_2O_2$) | −18.62 ($C_2$) |
| M06L | −1.16 | **4.26** | 0.21 | **4.56** | 17.72 ($P_4$) | −14.67 ($AlF_3$) |
| TM | 1.78 | 6.67 | 3.11 | 7.43 | 32.67 ($F_2O_2$) | −19.00 ($AlF_3$) |
| Hybrid | | | | | | |
| B3LYP | −2.22 | 3.40 | −2.94 | 4.28 | 6.37 ($N_2H$) | −31.87 (BN $^3\Pi$) |
| B3PW91 | −1.70 | 3.11 | −1.56 | 3.22 | 6.47 ($NO_2$) | −25.56 ($C_2$) |
| PBE0 | −2.47 | 3.75 | −2.47 | 3.99 | 5.68 ($NO_2$) | −31.04 (BN $^3\Pi$) |
| TPSSh | −1.73 | 4.58 | −1.78 | 4.90 | 15.17 ($B_2H_6$) | −34.73 (BN $^3\Pi$) |

### 3.1.3 Electron affinities

The electron affinity (EA) is the energy released when a free electron becomes attached to an atom or molecule. EA is defined as the difference between the total energies (including

ZPE) at 0 K of the neutral species and the corresponding anion. Listed in Table III are EA results calculated using TM along with those[19] of other functionals.

Table III. Summary of deviations from experiment for EAs of the G3/99 (58 species) test set. All values other than those of TM are from Ref. 19. The molecular geometries, electronic and unscaled zero-point energies of both the neutral and anion species by TM were evaluated using the 6-311++G(3*df*,3*pd*) basis set. All values are in eV. For non-hybrid functionals, the smallest and largest MAEs are in bold blue and red, respectively.

| Method | ME | MAE | Max (+) | Max (-) |
|---|---|---|---|---|
| Non-hybrid | | | | |
| LSDA | 0.23 | **0.24** | 0.88 ($C_2$) | -0.15 ($NO_2$) |
| BLYP | 0.01 | **0.12** | 0.70 ($C_2$) | -0.26 (NCO) |
| BPW91 | 0.04 | **0.12** | 0.78 ($C_2$) | -0.31 ($NO_2$) |
| BP86 | 0.18 | 0.19 | 0.89 ($C_2$) | -0.15 ($NO_2$) |
| PW91 | 0.11 | 0.14 | 0.84 ($C_2$) | -0.21 ($NO_2$) |
| PBE | 0.06 | **0.12** | 0.78 ($C_2$) | -0.29 ($NO_2$) |
| HCTH | 0.15 | 0.19 | 0.90 ($C_2$) | -0.27 (PH) |
| OLYP | -0.12 | 0.15 | 0.60 ($C_2$) | -0.47 ($NO_2$) |
| VSXC | -0.02 | 0.13 | 0.78 ($C_2$) | -0.35 ($NO_2$) |
| TPSS | -0.02 | 0.14 | 0.82 ($C_2$) | -0.32 ($NO_2$) |
| TM | -0.12 | 0.18 | 0.74 ($C_2$) | -0.45 (HOO) |
| Hybrid | | | | |
| B3LYP | 0.09 | 0.12 | 1.10 ($C_2$) | -0.09 (HOO) |
| B3PW91 | 0.03 | 0.14 | 1.08 ($C_2$) | -0.26 (HOO) |
| B3P86 | 0.59 | 0.59 | 1.63 ($C_2$) | None |
| PBE0 | -0.03 | 0.17 | 1.09 ($C_2$) | -0.39 (HOO) |
| TPSSh | -0.05 | 0.16 | 0.95 ($C_2$) | -0.33 (HOO) |

As seen from Table III, the TM functional underestimates EAs, similar to the other meta-GGAs listed. The MAE of TM (MAE=0.18 eV) is larger than the MAEs of many other functionals including TPSS and PBE, but smaller than the MAEs of HCTH, BP86, and LSDA. Generally, anions are artificially stabilized in finite-basis-set calculations.[65] Therefore, the comparison of errors may not be an explicit indication of the accuracy of the functional itself. The largest error of TM is found for the C$_2$ molecule, as for all other functionals, due to the multireference character of the singlet ground state of this molecule.[66–68]

Table IV. Summary of deviations from experiments of PAs for the G3/99 (8 species) test set. All values other than those of TM are from Ref. 19. The geometries, electronic and unscaled zero-point energies of TM were evaluated using the 6-311++G(3*df*,3*pd*) basis set. All values are in eV. For non-hybrid functionals, the smallest and largest MAEs are in bold blue and red, respectively.

| Method | ME | MAE | Max (+) | Max (-) |
|---|---|---|---|---|
| Non-hybrid | | | | |
| LSDA | -5.9 | **5.9** | None | -10.6 (PH$_3$) |
| BLYP | -1.5 | 1.6 | 0.4 (C$_2$H$_2$) | -3.9 (H$_2$O) |
| BPW91 | 0.9 | 1.5 | 3.8 (C$_2$H$_2$) | -1.3 (PH$_3$) |
| BP86 | -0.5 | 1.3 | 2.4 (C$_2$H$_2$) | -2.9 (PH$_3$) |
| PW91 | -0.9 | 1.6 | 2.2 (C$_2$H$_2$) | -3.5 (PH$_3$) |
| PBE | -0.8 | 1.6 | 2.4 (C$_2$H$_2$) | -3.6 (PH$_3$) |
| HCTH | 1.9 | 1.9 | 5.3 (C$_2$H$_2$) | None |
| OLYP | 1.5 | 1.7 | 5.4 (C$_2$H$_2$) | -0.6 (H$_2$O) |
| VSXC | 1.0 | 1.6 | 5.0 (C$_2$H$_2$) | -1.5 (H$_2$) |
| TPSS | 1.7 | 1.8 | 4.4 (C$_2$H$_2$) | -0.5 (H$_2$O) |
| TM | 0.7 | **1.2** | 4.3 (C$_2$H$_2$) | -1.5 (H$_2$O) |
| Hybrid | | | | |
| B3LYP | -0.8 | 1.2 | 1.6 (C$_2$H$_2$) | -2.3 (H$_2$) |
| B3PW91 | 1.0 | 1.1 | 4.2 (C$_2$H$_2$) | -0.3 (SiH$_4$) |
| B3P86 | 0.5 | 1.0 | 3.5 (C$_2$H$_2$) | -0.9 (SiH$_4$) |
| PBE0 | 0.2 | 1.1 | 3.9 (C$_2$H$_2$) | -1.7 (SiH$_4$) |
| TPSSh | 1.8 | 1.8 | 4.8 (C$_2$H$_2$) | None |

### 3.1.5 Proton affinities

The proton affinity (PA) of species is a measure of its gas-phase basicity. PA is defined as the difference between the ground-state energies (including ZPE) of the neutral and protonated species. The PAs for the 8 species of the G3/99 test set are listed in Table IV. We see that the TM functional gives the most accurate proton energies among non-hybrid DFT methods considered. Its error is comparable to those of hybrid functionals which, however, come with a higher computational cost.

Table V. Summary of deviations (in Å) from experiments of bond lengths ($r_e$) for the T-96R (96 diatomic molecules) test set. These are calculated using the 6-311++G(3$df$,3$pd$) basis set. All values other than those of TM are from Ref. 19. Hartree-Fock values do not include Be$_2$ (unbound). LSDA values do not include $F_2^+$ and SF (fails to converge). For non-hybrid functionals, the smallest and largest MAEs are in bold blue and red, respectively.

| Method | ME | MAE | Max (+) | | Max (-) | |
|--------|------|------|------|------|------|------|
| Non-hybrid | | | | | | |
| LSDA   | 0.001 | 0.013 | 0.042 | (BN) | -0.094 | (Na$_2$) |
| BLYP   | 0.021 | **0.022** | 0.055 | (Al$_2$) | -0.032 | (Na$_2$) |
| BPW91  | 0.017 | 0.017 | 0.070 | (Li$_2$) | -0.007 | ($F_2^+$) |
| BP86   | 0.017 | 0.018 | 0.060 | (Li$_2$) | -0.006 | ($F_2^+$) |
| PW91   | 0.014 | 0.015 | 0.054 | (Li$_2$) | -0.016 | (Be$_2$) |
| PBE    | 0.015 | 0.016 | 0.055 | (Li$_2$) | -0.013 | (Be$_2$) |
| HCTH   | 0.009 | 0.015 | 0.086 | (Na$_2$) | -0.087 | (Si$_2$) |
| OLYP   | 0.017 | 0.018 | 0.103 | (Na$_2$) | -0.017 | ($F_2^+$) |
| VSXC   | 0.012 | 0.013 | 0.085 | (Na$_2$) | -0.023 | (P$_4$) |
| TPSS   | 0.014 | 0.014 | 0.078 | (Li$_2$) | -0.008 | (P$_4$) |
| TM     | 0.010 | **0.012** | 0.054 | (Li$_2$) | -0.086 | (Si$_2$) |
| Hybrid | | | | | | |
| B3LYP  | 0.005 | 0.010 | 0.041 | (Be$_2$) | -0.040 | (Na$_2$) |
| B3PW91 | 0.003 | 0.009 | 0.060 | (Li$_2$) | -0.042 | ($F_2^+$) |
| B3P86  | 0.000 | 0.008 | 0.038 | (Be$_2$) | -0.044 | ($F_2^+$) |
| PBE0   | -0.001 | 0.010 | 0.063 | (Be$_2$) | -0.052 | ($F_2^+$) |
| TPSSh  | 0.008 | 0.010 | 0.074 | (Li$_2$) | -0.026 | ($F_2^+$) |

**3.2 Bond Lengths**

To evaluate the accuracy of the TM functional with regard to equilibrium bond lengths ($r_e$), we adopted the T-96R test set[19] of 96 ground-state molecules consisting of 10 molecular cations and 86 neutral molecules. The latter includes 73 diatomic molecules consisting of atoms ranging from H to Cl and 13 symmetric polyatomic molecules, each of which is characterized by a single bond length. The experimental values of equilibrium internuclear distances are taken from Ref. 69 for $Be_2$, Ref. 70 for NaLi and cations, and Ref. 71 for the rest. Table V shows that TM provides the most accurate description for molecular bond lengths, compared to other non-hybrid DFT methods, while it is slightly less accurate than hybrid functionals.

### 3.3 Harmonic Vibrational Frequencies

The harmonic vibrational frequency ($\omega_e$) is the frequency of the idealized harmonic vibration of the molecule. To evaluate the accuracy of the TM functional for harmonic vibrational frequencies, we used the T-82F test set[19] of 82 ground-state diatomic molecules, which includes 69 neutral species consisting of first- and second-row elements and 13 cations. The experimental values are from Ref. 69 for $Be_2$, Ref. 70 for NaLi and cations, and Ref. 71 for the rest. As shown in Table VI, TM is the most accurate non-hybrid functional for harmonic frequencies. Like other non-hybrid functionals, TM also underestimates vibrational frequencies, while hybrid functionals tend to overestimate them.

Table VI. Summary of deviations of the calculated harmonic vibrational frequencies from experiment for the T-82F (82 diatomic molecules) test set. For TM, the geometries and harmonic vibrational frequencies are computed using the 6-311++G(3$df$,3$pd$) basis set. All values other than those of TM are from Ref. 19. Hartree-Fock values do not include $Be_2$

(unbound). LSDA values do not include $F_2^+$ (fails to converge). All values are in cm$^{-1}$. For non-hybrid functionals, smallest and largest MAEs are in bold blue and red, respectively.

| Method | ME | MAE | Max (+) | Max (-) |
|---|---|---|---|---|
| Non-hybrid | | | | |
| LSDA | -11.8 | 48.9 | 140.7 ($F_2$) | -227.7 ($H_2$) |
| BLYP | -51.1 | **55.2** | 66.9 ($Be_2$) | -224.3 ($HF^+$) |
| BPW91 | -32.6 | 41.4 | 72.1 ($Be_2$) | -161.7 ($HF^+$) |
| BP86 | -37.7 | 45.5 | 71.4 ($F_2^+$) | -180.4 ($HF^+$) |
| PW91 | -29.3 | 39.8 | 82.1 ($Be_2$) | -170.1 ($HF^+$) |
| PBE | -31.7 | 42.0 | 82.5 ($Be_2$) | -175.3 ($HF^+$) |
| HCTH | -14.6 | 39.9 | 115.7 ($O_2^+$) | -116.9 (MgH) |
| OLYP | -28.7 | 40.2 | 89.4 ($F_2^+$) | -123.7 ($OH^+$) |
| VSXC | -12.2 | 33.9 | 100.3 ($N_2^+$) | -162.1 (BeH) |
| TPSS | -18.7 | 30.4 | 81.2 ($F_2^+$) | -145.9 (HF) |
| TM | -13.5 | **29.7** | 91.4 ($F_2^+$) | -145.2 (HF) |
| Hybrid | | | | |
| B3LYP | 9.5 | 33.5 | 161.9 ($F_2^+$) | -99.2 ($HF^+$) |
| B3PW91 | 21.9 | 36.2 | 194.0 ($F_2^+$) | -51.6 ($HF^+$) |
| B3P86 | 26.9 | 37.0 | 201.0 ($F_2^+$) | -52.3 ($HF^+$) |
| PBE0 | 34.7 | 43.6 | 236.3 ($O_2^+$) | -36.2 (AlH) |
| TPSSh | 6.6 | 26.7 | 141.4 ($F_2^+$) | -78.0 (HF) |

### 3.4 Reaction barrier heights

Calculation of reaction barrier heights presents a great challenge to semilocal DFT due to the presence of stretched bonds in a transition state. To evaluate the performance of the TM density functional for reaction barrier heights, we adopted the BH76 test set [72] which includes 38 hydrogen transfer barrier heights and 38 non-hydrogen transfer barrier heights. Since calculated reaction barrier heights are highly sensitive to the basis set, the geometries of the reactants, transition states, and products were optimized at QCISD/MG3 level. Single-point calculations of total electronic energies (not including ZPE) were performed with the MG3S basis set. Listed in Table VII are the errors for reaction barrier height BH76 set. Barrier heights for individual reactions are available in Table S8 of the supplementary material.[48] Like other semilocal functionals, the TM functional tends to underestimate reaction barrier heights (Table VII). The maximum positive and negative deviations are 20.95 and -28.89 kcal/mol, respectively. The MAE of TM is 7.08 kcal/mol, larger than those of the VSXC (MAE=4.77 kcal/mol), M06L (MAE=4.1 kcal/mol), and the hybrid functionals, but smaller than those of the SCAN (MAE=7.7 kcal/mol), TPSS (MAE=8.17 kcal/mol), PBE (MAE=8.71 kcal/mol), LSDA (MAE=14.88 kcal/mol), and the rest of the non-hybrid functionals.

TABLE VII. Summary of deviations of the calculated reaction barrier heights from CCSD(T) values[72] for the BH76 test set. Results are taken from Ref. 73 for M06L, Ref. 20 for SCAN, and Ref. 72 for other functionals. All values are in kcal/mol. For non-hybrid functionals, the smallest and largest MAEs are in bold blue and red, respectively.

| Method | ME | MAE |
|---|---|---|
| Non-hybrid | | |
| LSDA | -14.78 | **14.88** |
| BLYP | -8.09 | 8.11 |
| BP86 | -8.74 | 8.81 |
| PBE | -8.66 | 8.71 |
| VSXC | -4.56 | 4.77 |
| TPSS | -8.14 | 8.17 |
| M06L | -3.9 | **4.1** |
| SCAN | -7.7 | 7.7 |
| TM | -7.08 | 7.08 |
| Hybrid | | |
| B3LYP | -4.15 | 4.28 |
| TPSSh | -6.28 | 6.32 |

Reaction barrier heights are generally predicted more accurately by nonlocal density functionals (e.g., hybrid functionals incorporating exact exchange) because such functionals exhibit a lower delocalization error [27] for species with stretched bonds (i.e., transition states). From Table VII, we can see a large reduction of errors in reaction barrier heights from nonhybrid to hybrid functionals (e.g., from PBE to PBE0 and from TPSS to TPSSh).

### 3.5 Hydrogen-Bonded Complexes

Hydrogen bonds are ubiquitous in biomolecular systems, so accurate description of hydrogen-bonded systems is critically important for applications of DFT in computational biochemistry. Wave function-based *ab initio* methods such as second-order Møller-Plesset perturbation theory and coupled-cluster methods with good basis sets are highly accurate in describing weak bonding, but they are computationally demanding, especially for complex biomolecules. Therefore, density functionals that can accurately predict properties of weakly-bonded systems are highly desired. In this work, we adopted the test set of Rabuck and Scuseria[74] which includes 5 nonionic pairs $(HF)_2$, $(HCl)_2$, $(H_2O)_2$, HF/HCN, and HF/$H_2O$, as well as 5 ionic ones $CN^-/H_2O$, $OH^-/H_2O$, $HCC^-/H_2O$, $H_3O^+/H_2O$, and $NH_4^+/H_2O$. Table VIII reports a statistical summary for a calculation of the 10 dissociation energies ($D_0$) and 11 H-bond lengths. As seen from Table VIII, TM yields the most accurate H-bond dissociation energies for all DFT methods, including hybrid functionals. It also gives the most accurate H-bond lengths for non-hybrid functionals. It is even more accurate than many hybrid DFT methods.

Table VIII. Summary of deviations of bond lengths (Å) and dissociation energies $D_0$ (kcal/mol) of 10 hydrogen-bonded complexes. All errors are relative to the MP2(full)/6-311++G(3*df*,3*pd*) values.[19] All values other than those of TM are from Ref. 19. For the TM values, the 6-

311++G(3df,3pd) basis set is used in the calculations of both the geometry and unscaled ZPE-included dissociation energies. Illustration of bond lengths is available in Fig. 1 of Ref. 74. For non-hybrid functionals, the smallest and largest MAEs are in bold blue and red, respectively.

| Method | $D_0$ (kcal/mol) | | Bond lengths (Å) | |
| --- | --- | --- | --- | --- |
| | ME | MAE | ME | MAE |
| Non-hybrid | | | | |
| LSDA | 5.8 | **5.8** | -0.127 | 0.147 |
| BLYP | -0.5 | 0.6 | 0.027 | 0.034 |
| BPW91 | -0.7 | 1.0 | 0.008 | 0.045 |
| BP86 | 0.1 | 0.8 | -0.014 | 0.040 |
| PW91 | 1.4 | 1.4 | -0.028 | 0.052 |
| PBE | 0.9 | 1.0 | -0.018 | 0.043 |
| HCTH | -0.9 | 0.9 | 0.078 | 0.084 |
| OLYP | -2.2 | 2.2 | 0.136 | **0.157** |
| VSXC | -1.0 | 1.3 | 0.071 | 0.116 |
| TPSS | 0.3 | 0.6 | -0.006 | 0.021 |
| TM | -0.1 | **0.3** | 0.014 | **0.017** |
| Hybrid | | | | |
| B3LYP | -0.3 | 0.4 | 0.017 | 0.017 |
| B3PW91 | -0.5 | 0.9 | 0.005 | 0.035 |
| B3P86 | 0.4 | 0.7 | -0.023 | 0.043 |
| PBE0 | 0.5 | 0.7 | -0.012 | 0.032 |
| TPSSh | 0.1 | 0.5 | -0.002 | 0.015 |

### 3.6 Atomic excitation energies

Accurate prediction of excitation energies presents a great challenge to semilocal DFT, even in the non-adiabatic regime.[75] In this work, we assess the TM functional on the lowest singlet excitation energies of 13 atoms using time-dependent DFT [76,77] within the adiabatic approximation [78,79] and the 6-311++G(3df,3pd) basis set. As seen from Table IX, the TM functional yields the most accurate atomic excitation energies with an MAE of 0.40 eV, an error which is smaller than those of the LSDA (MAE=0.46 eV), PBE (MAE=0.53 eV), TPSS (MAE=0.47 eV), and even hybrid functionals B3LYP (MAE=0.44 eV), PBE0 (MAE=0.47 eV), and TPSSh (MAE=0.47 eV). The TM functional tends to overestimate atomic excitation energies (ME=+0.13 eV), unlike the other non-hybrid functionals considered. The superior performance of the TM approximation in adiabatic time-dependent DFT makes this semilocal functional potentially useful for simulation of dynamical properties of materials, for which hybrid functionals may be impractical due to their higher computational cost.

TABLE IX. Summary of deviations of the calculated lowest-lying singlet atomic excitation energies from experiment. The experimental values are from Ref. 80. The results of TM functional is calculated using the 6-311++G(3$df$,3$pd$) basis set. Results of LSDA, PBE, TPSS, TPSSh, PBE0, and B3LYP are taken from Ref. 79, except for those of O and F, which are evaluated in the present work. All values are in eV. The smallest and largest MAEs are in bold blue and red, respectively.

| Atom | Transition | LSDA | PBE | TPSS | TPSSh | PBE0 | B3LYP | TM | Expt. |
|---|---|---|---|---|---|---|---|---|---|
| He | 1s→2s | 19.59 | 19.73 | 20.27 | 20.58 | 20.62 | 20.50 | 20.44 | 20.62 |
|  | 1s→2s | 22.99 | 23.41 | 24.04 | 24.23 | 24.05 | 23.95 | 23.98 | 21.22 |
| Li | 2s→2p | 1.98 | 1.98 | 1.99 | 1.97 | 1.95 | 1.98 | 2.00 | 1.85 |
|  | 2s→3s | 3.12 | 3.09 | 3.09 | 3.13 | 3.23 | 3.16 | 3.22 | 3.37 |
| Be | 2s→2p | 4.84 | 4.91 | 5.06 | 5.05 | 4.94 | 4.88 | 5.01 | 5.28 |
|  | 2s→3s | 6.11 | 6.12 | 6.29 | 6.35 | 6.32 | 6.21 | 6.36 | 6.78 |
| Ne | 2p→3s | 17.45 | 17.21 | 17.55 | 17.94 | 18.27 | 17.88 | 17.76 | 16.62 |
|  | 2p→3p | 19.82 | 19.46 | 19.74 | 20.16 | 20.59 | 20.11 | 20.05 | 18.38 |
| Na | 3s→3p | 2.25 | 2.12 | 2.02 | 2.02 | 2.08 | 2.23 | 2.15 | 2.10 |
|  | 3s→4s | 3.05 | 2.91 | 2.87 | 2.90 | 3.02 | 3.02 | 3.05 | 3.19 |
| Mg | 3s→3p | 4.24 | 4.18 | 4.18 | 4.19 | 4.20 | 4.23 | 4.28 | 4.35 |
|  | 3s→4s | 5.02 | 4.93 | 5.01 | 5.06 | 5.08 | 5.00 | 5.12 | 5.39 |
| Ar | 3p→4s | 11.32 | 11.27 | 11.59 | 11.81 | 11.90 | 11.56 | 11.78 | 11.55 |
|  | 3p→4p | 12.68 | 12.50 | 12.74 | 13.00 | 13.22 | 12.89 | 12.98 | 12.91 |
| K | 4s→4p | 1.70 | 1.50 | 1.36 | 1.36 | 1.45 | 1.64 | 1.48 | 1.61 |
|  | 4s→5s | 2.52 | 2.35 | 2.28 | 2.30 | 2.42 | 2.43 | 2.43 | 2.61 |
| Ca | 4s→3d | 1.88 | 1.88 | 1.87 | 2.02 | 2.24 | 2.16 | 2.11 | 2.71 |
|  | 4s→4p | 3.09 | 2.98 | 2.90 | 2.90 | 2.96 | 3.03 | 3.01 | 2.93 |
| Zn | 4s→4p | 5.80 | 5.67 | 5.59 | 5.52 | 5.51 | 5.65 | 5.70 | 5.80 |
|  | 2s→5s | 6.38 | 6.12 | 6.10 | 6.12 | 6.20 | 6.22 | 6.31 | 6.92 |
| Kr | 4p→5s | 9.52 | 9.43 | 9.72 | 9.92 | 10.01 | 9.69 | 9.94 | 9.92 |
|  | 4p→5p | 10.84 | 10.64 | 10.85 | 11.10 | 11.30 | 10.98 | 11.09 | 11.30 |
| O | 2s→2p | 15.20 | 15.17 | 15.83 | 15.89 | 15.47 | 15.61 | 15.66 | 15.66 |
| F | 2s→2p | 19.51 | 19.89 | 20.65 | 20.90 | 20.70 | 20.57 | 20.89 | 20.90 |
| ME |  | -0.03 | -0.19 | -0.02 | 0.10 | 0.16 | 0.07 | 0.13 |  |
| MAE |  | 0.46 | **0.53** | 0.47 | 0.47 | 0.47 | 0.44 | **0.40** |  |

## 4. CONCLUDING REMARKS

In conclusion, we have made a comprehensive assessment of the nonempirical TM meta-GGA functional on standard molecular test sets. Our calculations show that, among all the non-hybrid functionals considered, the TM functional achieves consistently high accuracy for most properties. For excitation energies, proton affinities, harmonic vibrational frequencies, as well as dissociation energies and bond lengths of hydrogen-bonded complexes, it is competitive with or even more accurate than commonly used hybrid functionals, but has a lower computational cost, making the TM approximation an attractive candidate for molecular electronic structure calculations. This accuracy greatly benefits from the improved description of short-range interaction.

A striking feature of the TM functional is that it incorporates many exact constraints through the underlying exchange hole: (1) negativity, (2) uniform coordinate scaling,[41] (3) spin scaling relationship,[42] and (4) correct uniform-gas limit. These conditions are also satisfied by the density matrix expansion-based VSXC and M06-L meta-GGA functionals. The small-$u$ behavior of the exchange hole [81] (where $u$ is the separation between an electron and the hole around the electron) and the sum rule for the exchange hole are also incorporated into the TM functional. However, the exact fourth-order gradient expansion constraint has to be imposed separately, because the hole is only ensured to be correct in the uniform-gas limit.

The high accuracy of the TM functional greatly benefits from the fact that its exchange enhancement factor shows a slight oscillatory behavior,[82] like VSXC and M06-L. This behavior enables the TM functional to capture or extend the short-range part of the van der Waals interaction, due to the de-enhancement (relative to the LSDA) in some regions, leading to the improvement of noncovalent interactions, as demonstrated with hydrogen-bonded complexes (Table VIII) and molecular dimers.[83] However, since VSXC and M06 do not have the correct gradient expansion in the slowly varying limit and since they were trained only on molecular data sets, they are much more popular in quantum chemistry than in condensed-matter physics. Nevertheless, due to the recovery of the correct uniform-gas limit, those two functionals also perform quite well for solids. Unlike VSXC and M06L, the TM correlation functional was developed separately from the exchange part. It respects all the exact conditions that the TPSS correlation satisfies, and is an improvement over TPSS in the low-density (strong-interaction) limit.

Finally, it is worth pointing out that in the development of the TM functional, the Lieb-Oxford bound [84,85] has not been considered. The reason is that while this bound is an exact constraint for the integrated exchange energy,[86] it is locally violated by the conventional exact exchange energy density.[87]

In another paper [88] submitted elsewhere, we assess the performance of the TM functional for solids. Our results show that the TM functional also performs very well for periodic systems.

In particular, it yields the best lattice constants among many accurate density functionals included for comparison.

## SUPPLEMENTARY MATERIAL

See supplementary material for all calculated properties of individual species.

## ACKNOWLEDGMENTS

JT thanks John P. Perdew for useful discussions and suggestions. JT and YM acknowledge support from the NSF under Grant No. CHE 1640584. JT also acknowledges support from Temple University. GT was supported by China Scholarship Council and the National Natural Science Foundation of China under Grant No. 51264021. VNS was supported by the Natural Sciences and Engineering Research Council of Canada (NSERC). The work at Rice University was supported by the U.S. Department of Energy, Office of Basic Energy Sciences, Computational and Theoretical Chemistry Program under Award No. DE-FG02-09ER16053. GES is a Welch Foundation Chair (C-0036). Computational support was provided by the HPC of Temple University.

* Corresponding author. jianmin.tao@temple.edu